\documentclass[9pt,dvips]{article}
\usepackage{amsmath}
\begin{document}

\title{Shower Center of Gravity and Hadronic Interaction Characteristics}
\author{Lev Kheyn}
\date{ }
\maketitle

\begin{center}

\vspace{-0.4cm}
{\sl Skobeltsyn Institute of Nuclear Physics, Moscow State University\\
119991 Moscow, Russia}\\
\end{center}

\boldmath

\oddsidemargin=-10pt
\hspace*{0cm}

\begin{abstract}

Equations for the center of gravity of the shower originated by high energy proton in the atmosphere are written and,
within certain simplifications, solved for the case of logarithmically decreasing interaction length
of hadrons in the air. Obtained expression provides transparent view of the way in which hadronic interaction
characteristics determine the longitudinal shower development. 

\end{abstract}

\section{Introduction}

Since long ago, numerous attempts have been undertaken to explicitely connect air shower longitudinal profile,
in particular the shower maximum depth, with hadronic interaction characteristics. 

Very approximate approaches were tried, from toy models to extensions of Heitler model for the
electromagnetic shower \cite{Heitler} to the hadronic shower \cite{Matthews1,Matthews2,Stanev} and all these
have proven to be of not big quantitative help. 

A direct way to establish the connection is use of the cascade theory. A problem is that the shower maximum is
an inconvenient quantity for treatment by cascade equations. Rather, a convenient quantity is the shower center
of gravity (CG). It differs from the shower maximum basically by a shift. Thus, with CG one can trace the
elongation rate of the shower maximum. Even better suited CG for studying the difference in the elongation rate
between shower simulation codes using different interaction models and for separating main factors in the
interaction properties determining this difference. 

In this work, equations are written for the center of gravity of the proton initiated shower 
and within certain simplifications an analytical solution is derived.

\section{Derivation} 

We consider dependence on energy of the shower center of gravity:

$$ \overline{X (E)} = \int\limits_0^\infty X \, N(X) \, dX / \int\limits_0^\infty N(X) \, dX =
\int\limits_0^\infty X \, N(X) \, dX / \, E $$

For convenience, further we measure depth in the radiation units, i.e pass to variable $t=X/X_0$. Also, we measure
energy in units of the critical energy $E_c$.\\

The proton primary is considered. In simplifying assumptions of Feynman scaling and neglecting production of
other than nucleons and pions particles and neglecting pion decay, the system of equations for the
nominator  ${\Phi=\int\limits_0^\infty t\,N(t)\,dt}$ looks:

\begin{gather*}
\hspace*{-0.5cm}
\Phi_N(E)= \int\limits_0^1 \frac{dn_{N{\rightarrow}N}}{dx}(x) \, \Phi_N(Ex) \, dx + 
           \int\limits_0^1 \frac{dn_{N{\rightarrow}\pi}}{dx}(x) \, \Phi_\pi(Ex) \, dx  \\
           + \; \int\limits_0^1 \frac{dn_{N{\rightarrow}0}}{dx}(x) \, \Phi_0(Ex) \, dx +
	   E \, \lambda_N(E) \\	
\Phi_\pi(E)= \int\limits_0^1 \frac{dn_{\pi{\rightarrow}\pi}}{dx}(x) \, \Phi_\pi(Ex) \, dx + 
            \int\limits_0^1 \frac{dn_{\pi{\rightarrow}0}}{dx}(x) \, \Phi_0(Ex) \, dx +
	   E \, \lambda_\pi(E) \\ 	 
\Phi_0(E)=2 \, \int\limits_0^1 \frac{dx}{x}\Phi_\gamma(Ex) \, dx  	 
\end{gather*}
Here N denotes nucleon, $\pi$ denotes charged pion, 0 denotes $\pi^0$, $\gamma$ denotes photon 
and $dn_{i{\rightarrow}j}/dx$ defines distribution over Feynman x of the secondary particle of type j produced by 
the primary particle of type i.\\
                                                                                  
For the center of gravity of the electromagnetic shower, ingnoring the Landau-Pomeranchuk effect, we can obtain:
$$ \hspace*{0cm}
\overline{ t_\gamma(E)} = \ln{E}+\delta $$

Exact value of $\delta$ is unimportant for our goals, in the approximation B of cascade theory it proves
to be close to $\delta$ = 1.7.\\	 

Accordingly
$$ \hspace*{0cm}
\Phi_\gamma(E)= E \left(\ln{E}+\delta \right) $$

Let's apply Mellin transform: 
$$ F(s)= \int\limits_0^\infty E^{-s-1} \, F(E) \, dE $$

We obtain:
\begin{gather*}
\Phi_N(s)= \Phi_N(s) \,   \int\limits_0^1 x^s \, \frac{dn_{N{\rightarrow}N}}{dx}(x) \, dx \,  + 
           \Phi_\pi(s) \, \int\limits_0^1 x^s \, \frac{dn_{N{\rightarrow}\pi}}{dx}(x) \, dx \\
 + \, \Phi_0(s) \,   \int\limits_0^1 x^s \, \frac{dn_{N{\rightarrow}0}}{dx}(x) \, dx \,  + \lambda_N(s+1) \\
\Phi_\pi(s)= \Phi_\pi(s) \, \int\limits_0^1 x^s \, \frac{dn_{N{\rightarrow}\pi}}{dx}(x) \, dx \, +             
\Phi_0(s) \,   \int\limits_0^1 x^s \, \frac{dn_{N{\rightarrow}0}}{dx}(x) \, dx \,  +\lambda_\pi(s+1) \\ 
\Phi_0(s)= \frac{2\Phi_\gamma(s)}{s+1} = \frac{2}{s+1} \left\{ \frac{\delta}{s-1} + \frac{1}{(s-1)^2} \right\} 	 
\end{gather*}

Solution of the system of equations is:
$$ \hspace*{0cm}
\Phi_N(s)= \frac {\Phi_0(s)}{1-f_{NN}(s)} \, \left[ \frac{f_{N\pi}(s) \, f_{\pi0}(s)}{1-f_{\pi\pi}(s)} + f_{N0}(s) \right] 
+ \, \frac{\lambda_N(s+1)}{1-f_{NN}(s)} + \frac{f_{N\pi}(s)}{1-f_{NN}(s)} \cdot \frac{\lambda_\pi(s+1)}{1-f_{\pi\pi}(s)} $$	 
            
\[ \text{Here \;\;\;\;} f_{ij}(s)=\int_0^1 x^s \, \frac{dn_{i{\rightarrow}j}}{dx}(x) \, dx \] \\
            
Let's assume logarithmic dependence of the interaction length on energy:

$$\hspace*{0cm} \lambda(E) = \lambda_0 \{1-\alpha \log(E)\} $$

In that case 
$$ \lambda(s+1)=\lambda_0 \left\{ \frac{1}{s-1}-\frac{\alpha}{(s-1)^2} \right\} $$ \\

Let's apply inverse Mellin transformation to the solution:
$$ \hspace*{0cm}
\Phi_N(E)= \frac {1}{2\pi i} \oint E^s\Phi_N(s)\, ds $$ 

$\Phi_N(s)$ is a sum of two terms with poles at s=1:

$$ \Phi_N(s)=\Phi_N^{(1)}(s)\frac{1}{s-1} + \Phi_N^{(2)}(s)\frac{1}{(s-1)^2}$$ 

$$ 
\Phi_N^{(1)}(s)=\frac {2\delta}{(s+1)(1-f_{NN}(s))} \, \left[ \frac{f_{N\pi}(s) \, f_{\pi0}(s)}{1-f_{\pi\pi}(s)} + f_{N0}(s) \right]  
 + \, \frac{\lambda_N^{(0)}}{1-f_{NN}(s)} + \frac{f_{N\pi}(s)}{1-f_{NN}(s)} \cdot \frac{\lambda_\pi^{(0)}}{1-f_{\pi\pi}(s)} $$	 

$$ 
\Phi_N^{(2)}(s)=\frac {2}{(s+1)(1-f_{NN}(s))} \, \left[ \frac{f_{N\pi}(s) \, f_{\pi0}(s)}{1-f_{\pi\pi}(s)} + f_{N0}(s) \right]  
 - \, \frac{\lambda_N^{(0)}\cdot\alpha_N}{1-f_{NN}(s)} - \frac{f_{N\pi}(s)}{1-f_{NN}(s)} \cdot \frac{\lambda_\pi^{(0)}\cdot\alpha_\pi}{1-f_{\pi\pi}(s)} $$ \\	 

The complex integral equals to sum of residues at these poles.
$$\frac {1}{2\pi i} \oint E^s\Phi_N(s)\, ds = \Sigma res\, \Phi_N$$ 

The contribution of the first pole is
$$ 
\Phi_N^{(1)}(s=1) = E \left\{ \delta+\frac{1}{1-g_{NN}}\left( 
\lambda_N + \frac{g_{N\pi}}{g_{\pi0}} \, \lambda_\pi \right) \right\},	
$$            
where $  g_{ij} = f_{ij}(s=1) $ and we have taken into account that  
$g_{NN}+g_{N\pi}+g_{N0}=1$ and $g_{\pi\pi}+g_{\pi0}=1$. \\

The contribution of the second pole is 
$$\hspace*{0cm}
\lim_{s\to1} \, \frac{d}{ds} \left\{E^s \Phi_N^{(2)}(s) \right\},  $$ 
which equals to 
\begin{multline*}
E \left\{ \ln{E}-\frac{1}{2} + \frac{f_{NN}'+f_{N\pi}'+f_{N0}'}{1-f_{NN}} + 
 \frac{f_{N\pi}}{1-f_{NN}} \cdot \frac{f_{\pi0}'+ f_{\pi\pi}'}{1-f_{\pi\pi}} \right.\\  
\left. + \frac{\alpha_N \lambda_N^{(0)}}{1-g_{NN}} \left( y+\frac{f_{NN}'}{1-g_{NN}} \right) 
+ \frac{\alpha_\pi \lambda_\pi^{(0)}}{g_{\pi0}} \cdot \frac{g_{N\pi}}{1-g_{NN}} 
\left( y+\frac{f_{NN}'}{1-g_{NN}} +\frac{f_{N\pi}'}{g_{N\pi}}+ \frac{f_{\pi\pi}'}{g_{\pi0}} \right) \right\}, 
\end{multline*}
where 
$$f_{ij}' = \frac{d}{ds}f_{ij}(s)\left|_{s=1} 
= \int\limits_0^1 \, x \ln{x} \, \frac{dn_{i{\rightarrow}j}}{dx}(x) \, dx \right .$$

Let's denote 
\begin{multline*}
\mu_N=f_{NN}'+f_{N\pi}'+f_{N0}'= \int\limits_0^1 x \ln{x} 
\left\{\frac{dn_{N{\rightarrow}N}}{dx}(x)+\frac{dn_{N{\rightarrow}\pi}}{dx}(x)+\frac{dn_{N{\rightarrow}0}}{dx}(x) \right\} dx \\
= \; \int\limits_0^1 \, x \ln{x}\frac{dn_{N{\rightarrow}X}}{dx}(x) \, dx
\end{multline*}
and 
$$\mu_\pi=f_{NN}'+f_{\pi\pi}'+f_{\pi0}'= \int\limits_0^1 dx \, x \ln{x}  
\left\{\frac{dn_{\pi{\rightarrow}\pi}}{dx}(x)+\frac{dn_{\pi{\rightarrow}0}}{dx}(x) \right\} dx \\
= \; \int\limits_0^1 \, x \ln{x}\frac{dn_{\pi{\rightarrow}X}}{dx}(x) \, dx,$$ 
where index X in the inclusive distributions implies production of particles of any kind.\\

We obtain for the center of gravity
\begin{multline*}
\overline{t_N (E)}= \frac{1}{1-g_{NN}} \left\{ \lambda_N^{(0)} \left[1-\alpha_N \left(\ln{E}+\frac{f_{NN}'}{1-g_{NN}} \right) \right] + \mu_N  \right\} \\
+ \frac{g_{N\pi}}{(1-g_{NN}) \, g_{\pi0}} \left\{\lambda_\pi^{(0)} \left[1- \alpha_\pi \left(\ln{E}+\frac{f_{NN}'}{1-g_{NN}} +
\frac{f_{N\pi}'}{g_{N\pi}}+ \frac{f_{\pi\pi}'}{g_{\pi0}} \right)\right] + \mu_\pi \right\} + \ln{E}+\delta-\frac{1}{2}
\end{multline*} \\

This is equivalent to the interaction lengths being taken at some effective, reduced relative to the primary ones, energies:

$$E_N^{eff}=E / \exp{\left(\frac{f_{NN}'}{1-g_{NN}}\right)} \text{\;\;\;\;  and  \;\;\;\;} 
E_\pi^{eff}=E_N^{eff} / \exp{\left(\frac{f_{N\pi}'}{g_{N\pi}}+ \frac{f_{\pi\pi}'}{g_{\pi0}}\right)}$$\\

Finally, moving back to depth in $g/cm^2$ and energy in GeV:
$$
\overline{X_N(E)}= X_0 \left( \ln{\frac{E}{E_c}}+\delta-\frac{1}{2} \right)
+ \, \frac{1}{1-g_{NN}} \left\{ \lambda_N(E_N^{eff}) + X_0 \cdot \mu_N 
+ \, \frac{g_{N\pi}}{g_{\pi0}} \left[ \lambda_\pi(E_\pi^{eff}) + X_0 \cdot \mu_\pi \right] \right\} 
$$

$$\text{Interaction lengths are expressed trough inelastic cross-sections as }
\lambda= \frac{A}{N_A \, \sigma_{inel}}  \text{, \;  where A is atomic mass} $$ 
 and $N_A$ is Avogadro number. 

\section{Discussion} 

The final expression for the shower center of gravity explicitly splits into two terms: the center of gravity of the
purely electromagnetic cascade at primary energy and a modification of this by the hadronic cascade. The latter is
determined by two competing oppositely directed processes: i) carrying through energy by hadrons, that elongates the
shower, ii) energy dissipation in the hadronic interactions due to which electromagnetic subshowers start at smaller
energies, and because of the logarithmic energy dependency of their center of gravity that results in shortening of
the total shower. The first proccess is represented by $\lambda$ terms in the final expression, the second one is
represented by $\mu$ terms. 

Three main quantity characterizing hadronic interactions are usually considered as governing longitudinal shower
development: inelastic cross-sections, inelasticities and mean multiplicities. Cross-sections directly enter the
final expression for CG through interaction lenghs $\lambda$. Inelasticity, or more generally relative energy
transfers between different particle types, also directly enter the expression as $g_{ij}$ integrals (inelasticity
is the denominator in front of the curly braces). Whereas multiplicity does not enter the expression as such.
Instead, energy splitting is represented by the integral over inclusive distribution with additional weight
$x\ln{x}$ relative to the integral for the multiplicity. The meaning of this weight is clear: each produced
particle contributes to $\Phi$ with the weight which is the product of its energy and of the center of gravity,
which is proportional to the logarithm of this energy.

The obtained expression can provide only semi-quantitative results because of made simplifications, most  important
of which are neglecting Feynman scaling violation, production other than nucleons and pions particles and decay of
charged pions. Nevertheless, difference in the elongation rate predicted by different generators should not be
severely influenced by these simplifications.


\section{Conclusions} 

Equations for the center of gravity of the shower from primary high energy proton in the atmosphere are written and,
within  assumptions of Feynman scaling, neglecting production of other than nucleons and pions particles and
neglecting charged pion decay, solved for the case of logarithmically decreasing interaction length of nucleons and
pions. Hadronic interactions are represented by inelastic cross-sections, taken at some effective energies, and
integrals over inclusive distributions of two types: relative energy transfers between different particle types, like
inelasticity, defining elongation of the shower via hadronic cascading, and integrals with specific weight $x\ln{x}$,
reflecting energy dissipation and leading to shortening  of the total shower.

Results of applying the obtained expression will be presented in next publications.

\end{document}